\documentclass[]{AO4ELT}  
\pdfoutput = 1
\usepackage{microtype}
\usepackage{biblatex}
\usepackage{amsmath,amsfonts,amssymb}
\usepackage{graphicx}
\usepackage{pst-all} 
\usepackage[colorlinks=true, allcolors=blue]{hyperref}
\graphicspath{{./}}
\makeatletter         
\def\@maketitle{
\includegraphics[width = 170mm]{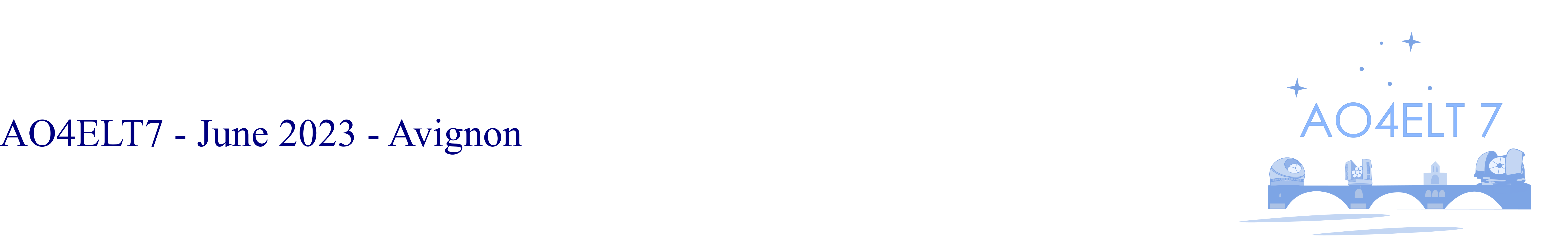}\\[8ex]
\begin{center}
{\Huge \bfseries \sffamily \@title }\\[4ex] 
{\Large  \@author}\\[4ex] 
\@date
\end{center}}

\title{Calibration Unit Design for High-Resolution Infrared Spectrograph for Exoplanet Characterization (HISPEC)}

\author[a]{Ben Sappey}
\author[a]{Quinn Konopacky}
\author[b]{Nemanja Jovanovic}
\author[b]{Ashley Baker}
\author[a]{Jerome Maire}
\author[b,c]{Samuel Halverson}
\author[b,c]{Dimitri Mawet}
\author[a]{Jean-Baptiste Ruffio}
\author[b]{Rob Bertz}
\author[d]{Michael Fitzgerald}
\author[b,c,e]{Charles Beichman}
\author[c]{Garreth Ruane}
\author[f]{Marc Kassis}
\author[d]{Chris Johnson}
\author[d]{Ken Magnone}
\author{HISPEC Team}

\affil[a]{Center for Astronomy and Astrophysics, University of California, San Diego, 9500 Gilman Dr. San Diego CA USA 92092}
\affil[b]{Department of Astronomy, California Institute of Technology, Pasadena, CA 91125, USA}
\affil[c]{Jet Propulsion Laboratory, California Institute of Technology, 4800 Oak Grove Drive, Pasadena, CA 91109, USA}
\affil[d]{Department of Physics \& Astronomy, 430 Portola Plaza, University of California, Los Angeles, CA 90095, USA}
\affil[e]{NASA Exoplanet Science Institute, California Institute of Technology, 770 South Wilson Avenue, Pasadena, CA 91125, USA}
\affil[f]{W.M. Keck Observatory, 65-1120 Mamalahoa Highway, Kamuela, HI, USA}

\authorinfo{Further author information: (Send correspondence to B.S.)\\B.S.: E-mail: bsappey@ucsd.edu}

\begin{document} 
\maketitle

\begin{abstract}
\par The latest generation of high-resolution spectrograph instruments on 10m-class telescopes continue to pursue challenging science cases. Consequently, ever more precise calibration methods are necessary to enable trail-blazing science methodology. We present the High-resolution Infrared SPectrograph for Exoplanet Characterization (HISPEC) Calibration Unit (CAL), designed to facilitate challenging science cases such as Doppler imaging of exoplanet atmospheres, precision radial velocity, and high-contrast high-resolution spectroscopy of nearby exoplanets. CAL builds upon the heritage from the pathfinder instrument Keck Planet Imager and Characterizer (KPIC) \cite{mawetKeckPlanetImager2016} \cite{delormeKeckPlanetImager2021} \cite{delormeFirstVersionFiber2018} and utilizes four near-infrared (NIR) light sources encoded with wavelength information that are coupled into single-mode fibers. They can be used synchronously during science observations or asynchronously during daytime calibrations. A hollow cathode lamp (HCL) and a series of gas absorption cells provide absolute calibration from 0.98 $\mu$m to 2.4 $\mu$m. A laser frequency comb (LFC) provides stable, time-independent wavelength information during observation and CAL implements a lower finesse astro-etalon as a backup for the LFC. Design lessons from instrumentation like HISPEC will serve to inform the requirements for similar instruments for the ELTs in the future.  
\end{abstract}

\keywords{High-resolution spectroscopy, astronomical instrumentation, calibration, precision radial velocity}

\section{INTRODUCTION}
\label{sec:intro}
\subsection{HISPEC Design Philosophy}

HISPEC is a high-resolution, diffraction-limited, simultaneous multi-band, fiber-fed spectrograph for W.M. Keck Observatory designed to be an ultra-stable instrument to achieve its science goals. The instrument will operate in the near-infrared (NIR) from 0.98 $\mu$m to 2.46 $\mu$m and will take spectra in \textit{y-}, \textit{J-}, \textit{H-}, and \textit{K-}band in a single exposure. HISPEC will be fed by the Keck adaptive optics (AO) system, either the visible Shack-Hartmann wavefront sensor (SHWFS) \cite{Wizinowich_2000} or the NIR pyramid wavefront sensor (PyWFS) \cite{bond_pywfs_2020} working in tandem with a deformable mirror (DM) to supply a diffraction-limited image to HISPEC’s input known as front-end instrument (FEI). The target light is coupled into single-mode optical fiber (SMF) where it is then routed to two spectrographs with Teledyne Hawaii H4RG detectors located in the basement of Keck for a further layer of thermal and vibrational isolation. This “light routing” is accomplished in a flexible way using mechanical fiber switchers that allow for optical fibers fed from different locations to be routed to the spectrograph. This allows the HISPEC instrument to be used in many different modes, e.g., a precision RV mode where the science target, background, and calibration light are all being taken simultaneously across the detector. The calibration system of HISPEC also contains a mechanical fiber switcher, allowing for more flexibility during daytime calibrations and reference calibration during science observations.

\subsection{CAL Design Overview}
HISPEC has challenging scientific goals that necessitate a carefully designed calibration system. The desired precision for both kinematic and relative flux measurements requires detailed calibration measurements of detector components and simultaneous tracking of drifts in wavelength and/or line spread function. The calibration subsystem (CAL) encompasses all components of HISPEC that result in calibration light reaching the spectrographs. It consists of a series of both relative and absolute calibration sources and the hardware that facilitates their routing to the rest of HISPEC.  

There are three major subdivisions for the CAL subsystem. The first is the absolute wavelength reference and flat-field unit, which utilizes lamps or gas cells with known atomic/molecular species to provide a stable reference point to "ground truth" wavelengths. The second is a Fabry-Perot etalon unit, provided by an external vendor, which offers precise relative wavelength calibration by referencing a well-known source. The third is the Laser Frequency Comb unit, which is being developed by both our team and an external vendor to provide what we hope will be the go-to relative wavelength solution reference for the majority of HISPEC science cases. Here, we describe the technical specifications and details of the CAL subsystem design.

\section{Calibration Methods}
\subsection{Dark Frames}
Dark frames are achieved by terminating all spectrograph input fibers with Narcissus mirrors (e.g., \href{https://www.thorlabs.com/thorproduct.cfm?partnumber=P5-1060R-P01-1}{Thorlabs P5-1060R-P01-1}). This means most of the light that is coupled into the fiber tips during dark frames is thermal radiation from inside the spectrograph itself, minimizing any light leaks or thermal instability. These Narcissus mirrors are integrated into the design in the CAL fiber switch system.
\subsection{Flat-fielding Detectors}
Understanding the pixel-to-pixel properties and variations across a Teledyne Hawaii H4RG detector is critical to getting to $<$1 m/s RV precision and as such the capability to flat-field the detectors often and repeatably is required. A good flat-field requires a flat distribution of photon flux across the detector to accurately measure variations in the detector pixels themselves. To this end, CAL provides the ability to flat-field leveraging the SMF design of HISPEC in a variety of scenarios, including controlling the flux density and spectral range projected on each detector. This is accomplished while maintaining the thermal integrity of the cryostat system by removing the light sources from the cryogenic environment where their heat would dissipate into the environment and disturb the thermal stability of the instrument. The flat-fielding system is divided in two: one unit to flat-field each detector. These solutions are identical in form but optimized for the respective wavelength ranges of BSPEC and RSPEC. Broadband infrared light is generated inside the CAL rack-mounted box by a tungsten-halogen lamp from Thorlabs (\href{https://www.thorlabs.com/thorproduct.cfm?partnumber=SLS202L}{Thorlabs SLS 202L}). This component was chosen due to its off-the-shelf availability, cost, and serviceability in addition to the adequate incident power produced by the source. While in free space, the light passes through a six-position filter wheel containing one of four bandpass filters and two open positions. The selected filters are shown in Table \ref{tab:blue} (BSPEC) and Table \ref{tab:red} (RSPEC).

\begin{table}[h]
    \centering
    \begin{tabular}{c|c|c|c|c}
       Vendor  & PN & Center Wavelength (nm) & FWHM Bandpass (nm) & Transmittance  \\
       \hline
       \hline
       Edmund Optics  & \#87-811 & 1000 & 25 & $>$90\% \\
       \hline
       Edmund Optics  & \#87-813 & 1100 & 25 & $>$90\% \\
       \hline
       Edmund Optics  & \#87-815 & 1200 & 25 & $>$90\% \\
       \hline
       Edmund Optics  & \#87-817 & 1300 & 25 & $>$90\% \\
       \hline
    \end{tabular}
    \caption{HISPEC CAL Filter Wheel Filter Selection BSPEC}
    \label{tab:blue}
\end{table}
\vspace{12pt}

\begin{table}[h]
    \centering
    \begin{tabular}{c|c|c|c|c}
       Vendor  & PN & Center Wavelength (nm) & FWHM Bandpass (nm) & Transmittance  \\
       \hline\hline
       Edmund Optics  & \#87-821 & 1500 & 25 & $>$90\% \\
       \hline
       Edmund Optics  & \#87-859 & 1600 & 50 & $>$90\% \\
       \hline
       Thorlabs  & FB1900-200 & 1900 & 200 $\pm$ 40 & $>$85\% \\
       \hline
       Thorlabs  & FB2300-50 & 2300 & 50 & $>$70\% \\
       \hline
    \end{tabular}
    \caption{HISPEC CAL flat-field Filter Wheel Filter Selection RSPEC}
    \label{tab:red}
\end{table}

\begin{figure}
    \centering
    \includegraphics[width=\linewidth]{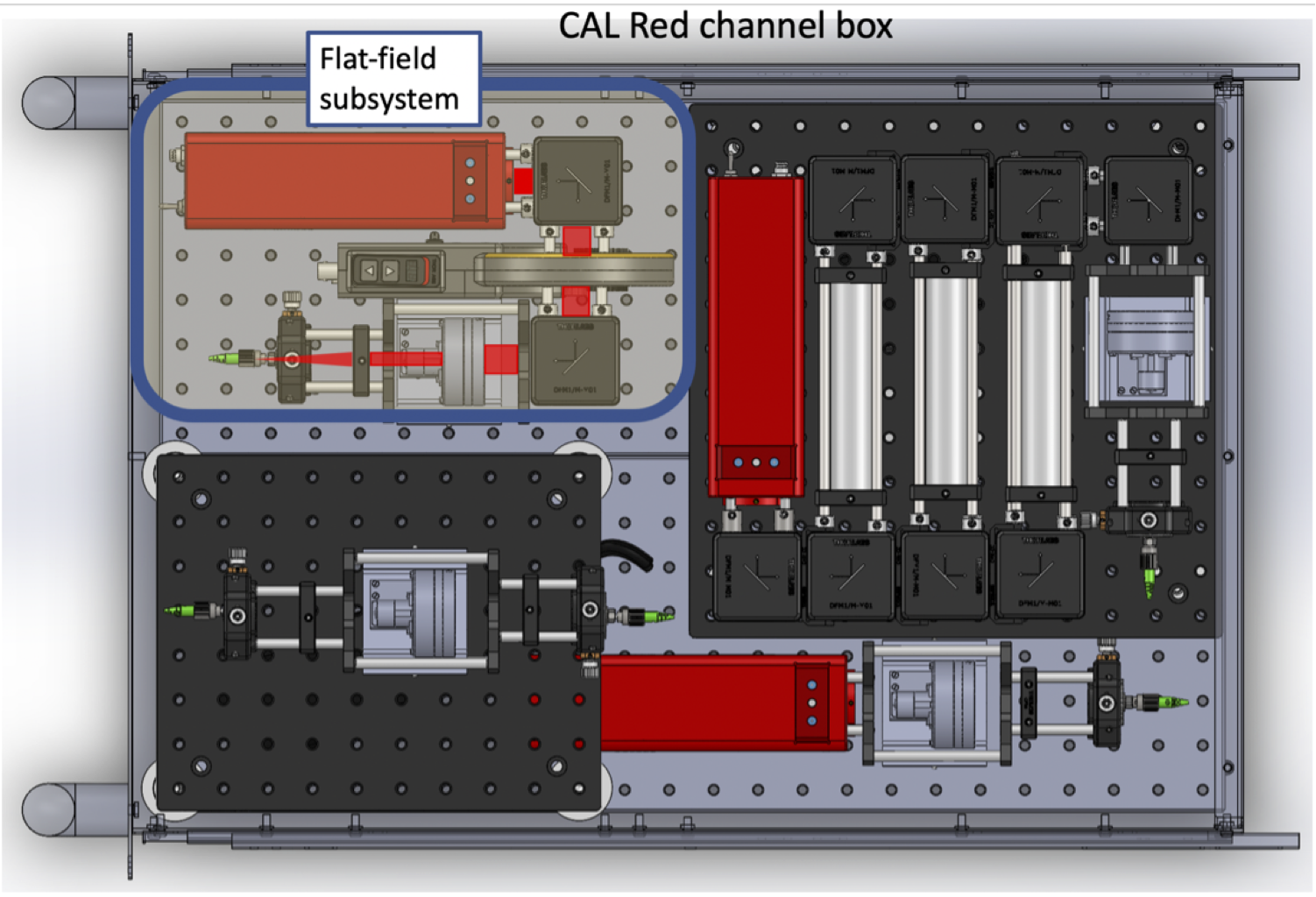}
    \caption{HISPEC RSPEC CAL Box with flat-fielding Unit Highlighted}
    \label{fig:RSPEC_CAL}
\end{figure}

\begin{figure}
    \centering
    \includegraphics[width = \linewidth]{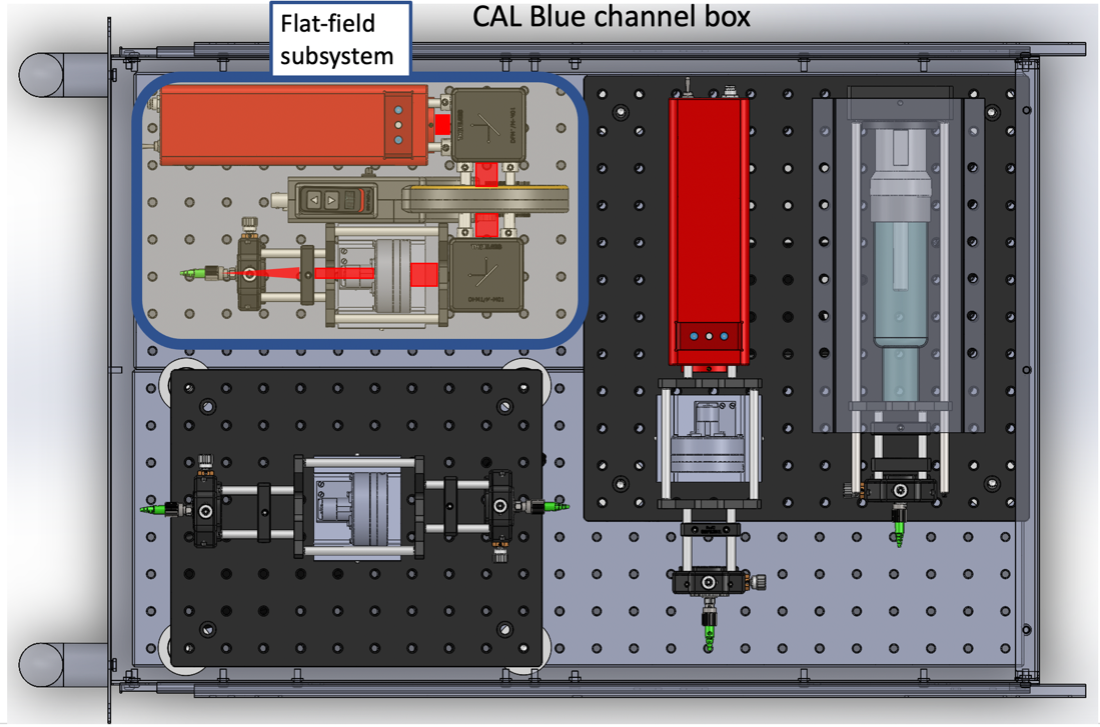}
    \caption{HISPEC BSPEC CAL Box with flat-fielding Unit Highlighted}
    \label{fig:BSPEC_CAL}
\end{figure}

\begin{figure}
    \centering
    \includegraphics{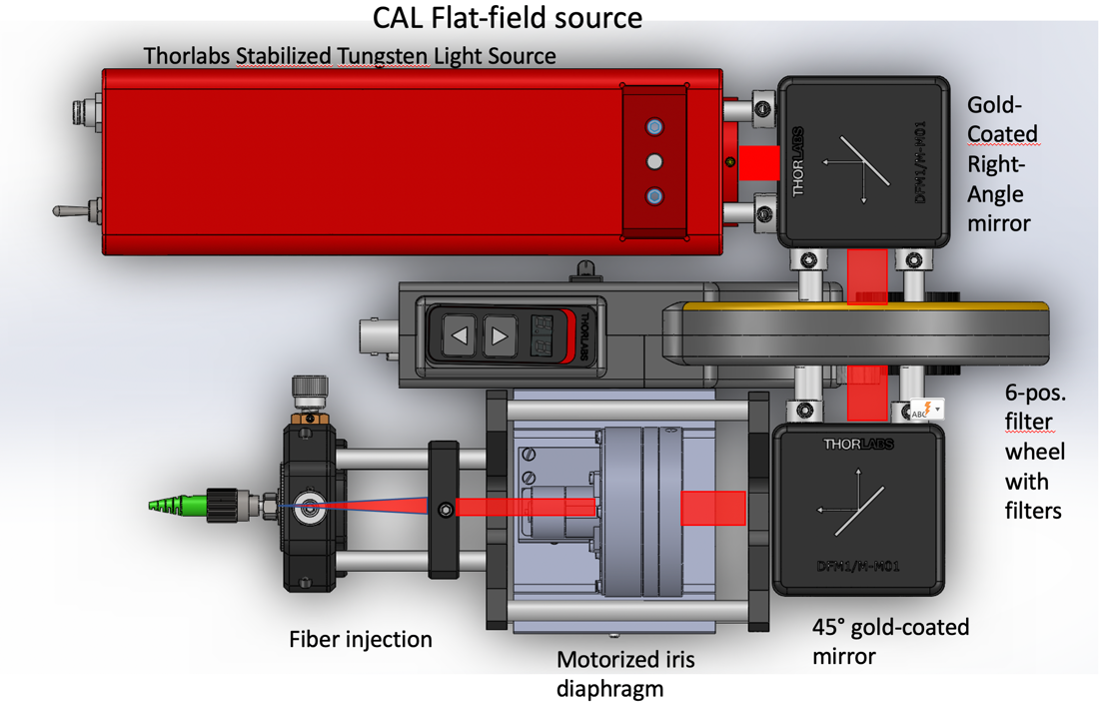}
    \caption{Light path of the Broadband NIR Source used for flat-fielding HISPEC}
    \label{fig:CALtopdown}
\end{figure}

Using bandpass filters affords the ability to measure the wavelength-dependent intra-pixel spectral response of the Teledyne Hawaii H4RG detectors while maintaining the classic near-blackbody tungsten-halogen lamp SED of the Thorlabs SLS 202L when the filter wheel is in the open position. An image of the CAL Red channel box and CAL Blue channel box  highlighting the flat-fielding system is shown in Figure \ref{fig:RSPEC_CAL} and \ref{fig:BSPEC_CAL} respectively. Figure \ref{fig:CALtopdown} shows the light path inside the flat-fielding subsystem.  Pseudo-collimated light exits the Thorlabs SLS 202L lamp where it is directed through a six-position filter wheel. In the open position, all the light passes through to the motorized iris diaphragm, which allows for an achromatic attenuation of the light before being focused and injected into a single mode fiber on an XY translation stage. Because the beam waist of the focused light is greater than the diameter of the single-mode fiber core, the XY translation adjustment is the only necessary adjustment. In the case where a spectral filter is inserted, the bandpass selected will be dispersed on the detector where that bandpass can be better characterized. These boxes will be rack-mounted in the basement of Keck.

\begin{figure}
    \centering
    \includegraphics{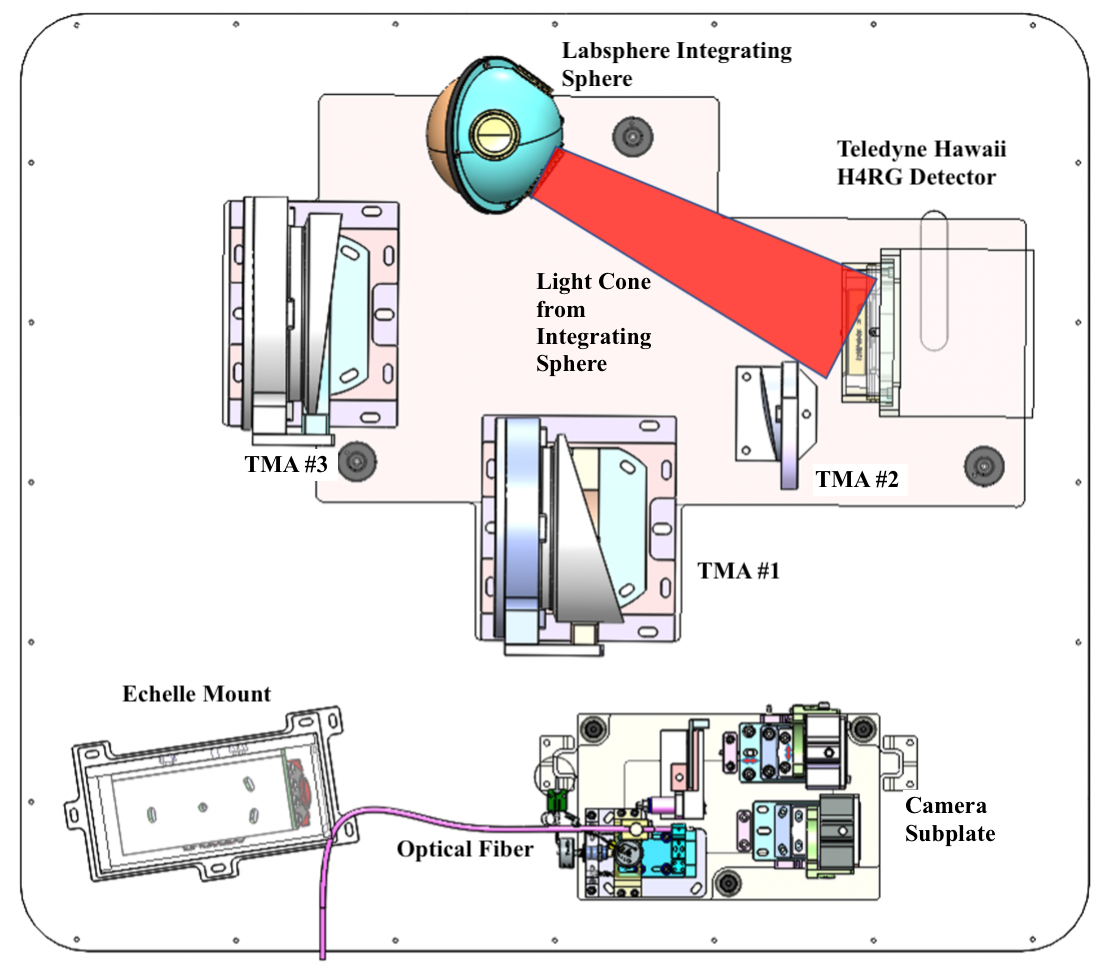}
    \caption{Top-Down view of the HISPEC RSPEC Spectrograph including the integrating sphere and flat-fielding light path}
    \label{fig:dewar}
\end{figure}

After being routed to the main body of the spectrograph, the light in the fiber is passed through the vacuum feedthrough specific to flat-fielding. Once inside the cryostat, the fiber is connected to an integrating sphere, supplied by Labsphere. We have selected the 3-port, 3-inch diameter Infragold-coated sphere for two reasons. First, this sphere has already been implemented in the Palomar Radial Velocity Instrument (PARVI) \cite{gibsonDataReductionPipeline2022}\cite{Cale_2023} to flat-field their Teledyne Hawaii H2RG detector at cryogenic temperature without issue and has been validated at similar wavelength ranges for use in a cryogenic environment. Second, the Infragold coating provides $\sim 96\%$ Lambertian reflectance in the NIR across our entire wavelength range, meaning that the flat-field will maintain the SED of the light source. At the output window of the integrating sphere, diffused light will travel a minimum of 13.3 cm to the surface of the detector. This distance will ensure a 1-2\% variation in the intensity of light across the diameter of the detector according to Labsphere. The notional light path of the flat-field inside of the cryostat is shown in Figure \ref{fig:dewar}.

\subsection{Absorption Gas Cells}
\label{sec:title}

\begin{figure}
    \centering
    \includegraphics[width=\linewidth]{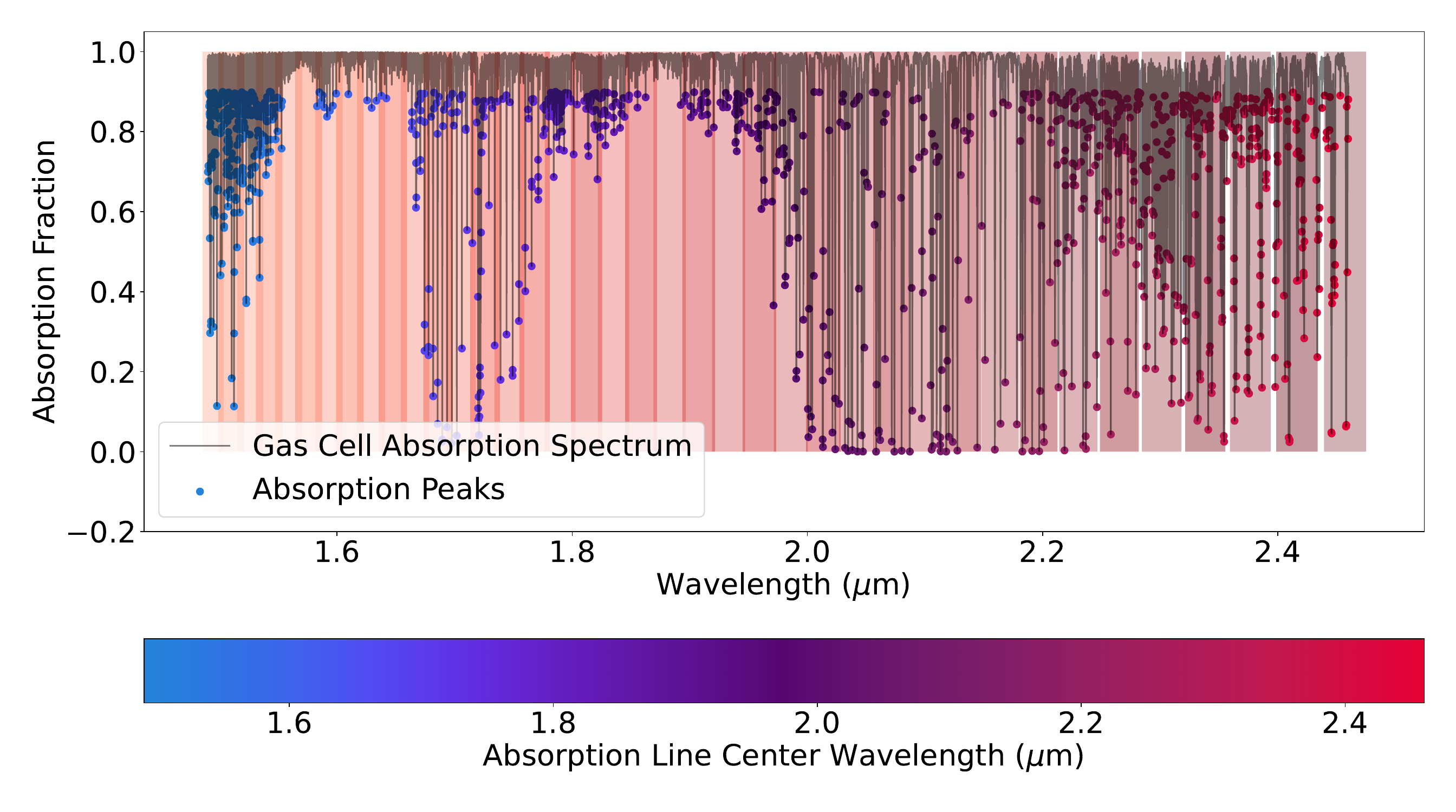}
    \caption{HISPEC CAL Gas Cell Absorption Combined Spectra with absorption line peaks $>10\%$ marked. The gases used are N$_2$O, NH$_3$, H$_2$O, CO$_2$, CH$_4$, and H$_2$S}
    \label{fig:gascellspectra}
\end{figure}

Understanding the wavelength calibration of the spectrograph and how it changes with time is the other key aspect that needs to be carefully calibrated, including both relative and absolute calibration. To enable this, CAL will provide several light sources the observer can choose from. The simplest of these are the gas cells (GAS) and arc lamps (LMP). These are summarized in Table \ref{tab:gases}. Three optimized gas cells (with room for additional cells) for the red spectral channel of HISPEC will be used to offer the most spectral coverage and lines/order for calibration. Gases were selected by generating predicted spectra using HITRAN Application Program Interface (HAPI) \cite{kochanovHITRANApplicationProgramming2016} which fetches data from \textbf{HITRAN} and uses the spectral information to simulate a gas cell environment where the gases, partial pressures, and cell length are variable. The goal of this exercise was to select a combination of gases that: 1) contained at least two absorption features in each order, 2) contained features that absorbed more than 5\% of the incident light to ensure detectable signal above the noise, 3) had appropriate pressure broadening properties such that the central wavelength of the absorption peak could be measured, and 4) would be minimally reactive in combination, even during the manufacturing process when high temperatures are used to melt the glass cell to seal it. The corresponding theoretical spectra of the cells are shown in Figure \ref{fig:gascellspectra}.  We focus on optimizing the gas cells for the red spectral channel ($\sim1.4 - 2.6  \mu$m) as gases that absorb from 0.960 – 1.320 $\mu$m are not readily available to be manufactured whereas hollow cathode lamps that contain a high density of emission lines in this wavelength range are much more common. 

\begin{table}
    \centering
    \begin{tabular}{c|c|c|c|c}
        Cell Name & Gases & Pressure (torr) & Cell Length (cm) & For Channel \\
        \hline
        \hline
         R1 & N$_2$O, NH$_3$ & 150, 150 & 10 & Red\\
         R2 & H$_2$O, CO$_2$ & 225, 225 & 10 & Red\\
         R3 & CH$_4$, H$_2$S & 300, 300 & 20 & Red\\
         &&&&\\
         Lamp name & Emission Species & Fill Gas & & \\
         \hline
         L1 & Thorium & Argon & & Both \\
         L1 & Uranium & Neon & & Both \\
         \hline
    \end{tabular}
    \caption{Absolute Wavelength Calibration Species Selection for RSPEC and BSPEC}
    \label{tab:gases}
\end{table}

The three gas cells will be produced by Wavelength References, the same company used to make KPIC’s gas cells. Though gas cells can be modeled under ideal circumstances, the wavelengths of spectral lines from a given gas cell are not always well understood. This is because when gases are mixed, they can react by introducing new transitions, some of which are poorly documented, or contain contaminates despite best practices in manufacturing. We will, therefore, utilize an R$\sim$200,000 Fourier transform spectrograph at the Jet Propulsion Laboratory to provide a base truth spectrum of the gas cells. Then, once CAL is integrated and commissioned with the rest of HISPEC, we can use the base truth spectrum to calibrate what data we receive from the instrument. Shifts in the temperature of the gases causes could impact the resulting spectral features as observed on HISPEC. Although simulations have shown that temperature changes in the expected range (i.e., laboratory room temperature to Maunakea summit temperature) have minimal effect on line width at the resolving power of HISPEC, gas cells will be further temperature controlled using polyimide flexible heating adhesive tape. With this information provided by JPL, line-matching techniques that are used to calculate the wavelength solution of the instrument become much more precise.

\begin{figure}
    \centering
    \includegraphics[width=0.85\linewidth]{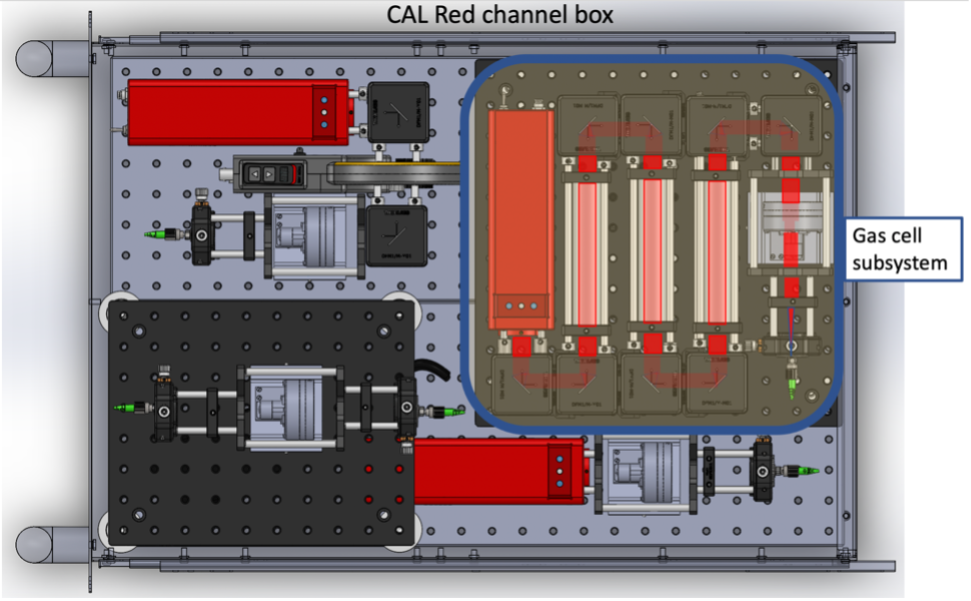}
    \caption{The gas cell subsystem is highlighted inside of the CAL red channel box. A Thorlabs SLS 202L broadband lamp is collimated and then directed through 3 gas cells, attenuated via a mechanical addressable iris, and then coupled into SMF before being routed to the spectrograph.}
    \label{fig:gas_cell}
\end{figure}

The gas cells will be in the CAL Red Box as shown in Figure \ref{fig:gas_cell}.

\subsection{Hollow Cathode Lamps}
\begin{figure}
    \centering
    \includegraphics[width=\linewidth]{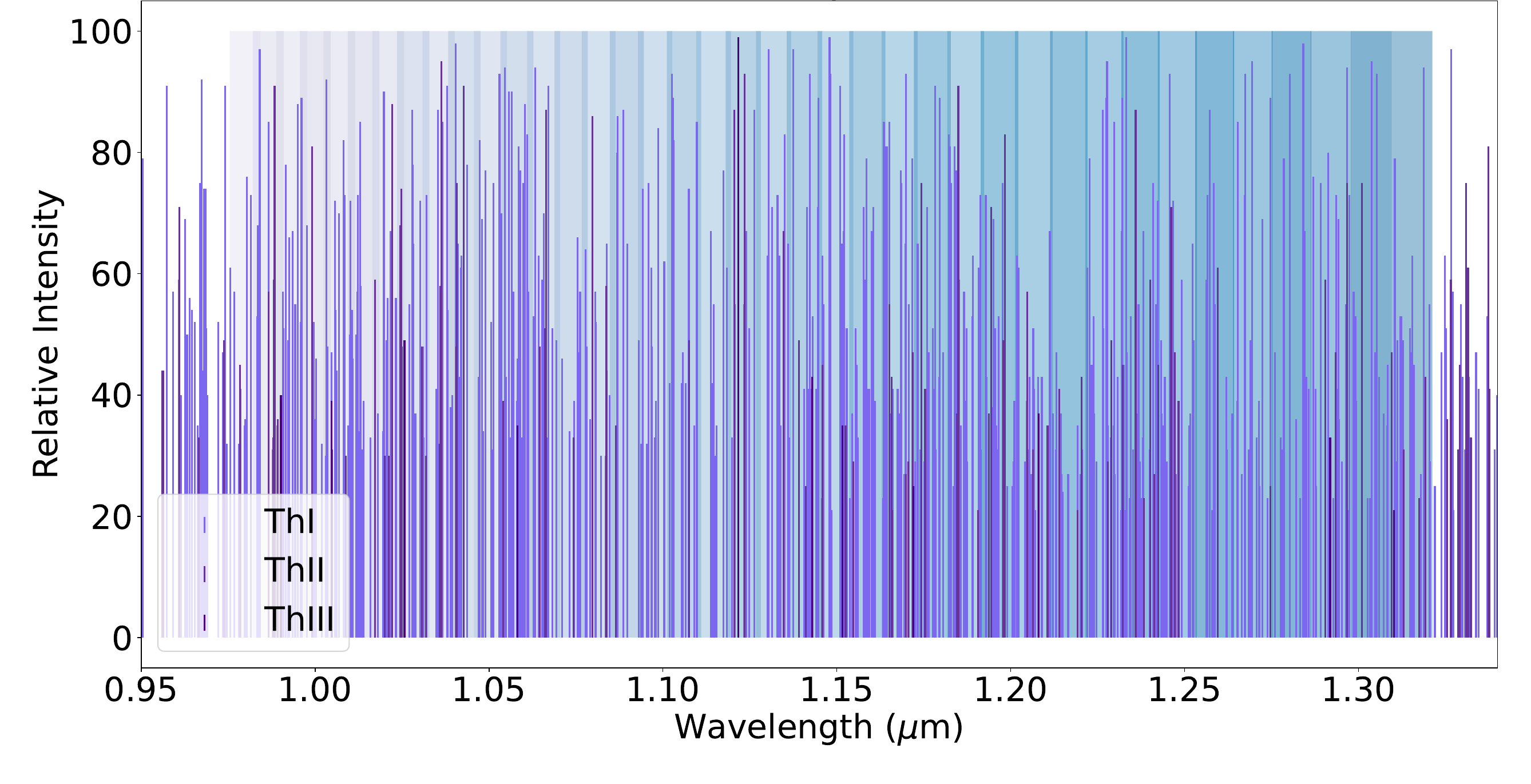}
    \caption{Thorium Emission lines plotted over BSPEC Orders. Each BSPEC Order is represented by the blue bars which form a gradient over the spectral bandpass. More than two emission lines are present in each BSPEC Order, meaning the entire order can be calibrated on the detector in an absolute sense.}
    \label{fig:emission}
\end{figure}

Given the lack of common NIR-absorbing gases in the BSPEC bandpass (0.98 - 1.327 $\mu$m), a different strategy must be used for absolute wavelength calibration. Historically this has been accomplished by using arc lamps or metal-halide hollow cathode lamps. Thorium/argon lamps are frequently used for PRV instrument calibrations; most recently they were been used to calibrate ESPRESSO at VLT in combination with a Fabry-Perot etalon \cite{Schmidt_2021_ESPRESSO}, though they have also been used to calibrate precision radial velocity instruments like MAROON-X \cite{MAROON-XINST} PRV spectrograph as well as European Southern Observatory's (ESO) High Accuracy Radial velocity Planet Searcher (HARPS) \cite{Cosentino_HARPS} in the past. However, recently produced thorium hollow cathode lamps have been shown to be contaminated with thorium oxide (ThO) \cite{Nave_ThO}, making it difficult to identify the stable thorium emission lines amongst a forest of moving ThO emission lines. Many recent high-PRV instruments implement a uranium/neon hollow cathode lamp for absolute wavelength calibration, including the Apache Point Observatory Galactic Evolution Experiment (APOGEE) Spectrographs \cite{2019PASP..131e5001W}, the CRyogenic high-resolution InfraRed Echelle Spectrograph (CRIRES+) \cite{Seemann_UNe_CRIRESplus}, and the Habitable Zone Planet Finder (HPF) \cite{2012SPIE.8446E..1SM}. In the optical band, U/Ne is an optimal choice for astronomy due to its high information density at this bandpass, though as we explore past 1.2 $\mu$m and certainly at 2.460 $\mu$m, that information density decreases. We therefore decided to investigate the use of a thorium/argon hollow cathode lamp which has been used previously for near-infrared astronomy and provides complete wavelength information about the blue spectra channel. Using the Th/Ar atlas in Redman et al. 2014 \cite{Redman_2014_Th}, which combines data sets (and cross validates them) from the literature, the emission line locations along with their relative intensities (given in arbitrary SNR) are determined for both the red and blue channels of HISPEC. The Th/Ar lamp light is coupled directly into SMF where it is routed out of the Blue CAL box to the CAL switch at which point it may be routed to BSPEC Calibration fiber when in use. The theoretical HCL emission line locations are shown in Figure \ref{fig:emission}. 
\subsection{Laser Frequency Combs}
An astrocomb is a Laser Frequency Comb (LFC) calibration source for astronomical spectrographs. Like gas cells, arc lamps, and etalons, it provides light at well-defined frequencies to serve as a “spectral ruler” to compare against target object spectral signatures. As such, it is an important tool for performing precision radial velocity studies of exoplanet-hosting stars. Unlike gas cells and hollow cathode lamps, LFCs provide light “markers” at very uniform spacing. Generally, the line spacing of an LFC is very close, but in the case of an astrocomb, the spacing must be optimized for a particular spectrograph’s resolution. Unlike etalons, the spacing between the lines is traceable to a GPS-disciplined clock signal, meaning the relative and absolute positions of the light markers are highly stable. Astrocombs are therefore the highest precision and accuracy calibration sources available to astronomical spectrographs. They are also the most complex of these instruments. 
 
There will be two astrocomb sources available to the HISPEC spectrograph; each generates calibration lines by a different method. The first of these is a commercially available astrocomb manufactured by Menlo Systems and supplied to the W.M. Keck Observatory for the Keck Planet Finder (KPF) instrument \cite{gibson_howard_roy_smith_2018}. This comb uses spectrally flattened “excess” infrared light produced from the KPF comb. It provides a comb spectrum with a 20 GHz line spacing from 0.970 $\mu$m through 1.450 $\mu$m with frequency stability performance of $<$10 cm/s. Menlo combs are mode-locked laser combs that are spectrally filtered through a series of Fabry-Perot filter cavities to achieve a spectrum sparse enough to be resolved on an astronomical spectrograph. 

\begin{figure}
    \centering
    \includegraphics[width=\linewidth]{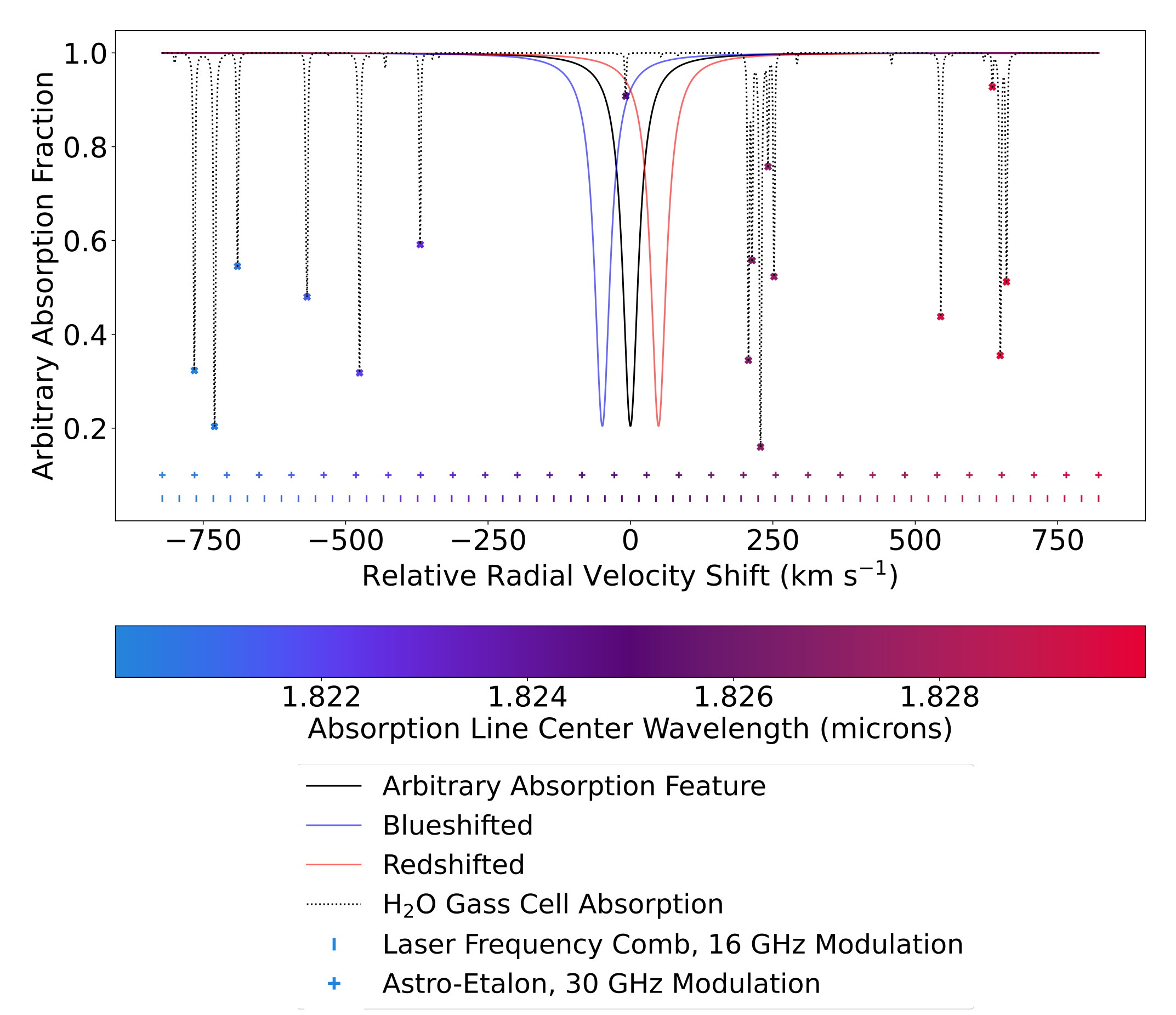}
    \caption{HISPEC Absolute and Relative Wavelength Calibration Methods. An arbitrary absorption feature is both red-shifted and blue-shifted which can be measured using the ruled marker of the \textit{h-, k-} astrocomb, modulated at 16 GHz, and the \textit{h-, k-} astro-etalon modulated at 30 GHz. The absorption lines of the H$_2$O gas cell are also shown for absolute reference}
    \label{fig:absrel}
\end{figure}

The second comb is an electro-optic (EO) LFC that generates calibration lines by creating sidebands on a continuous-wave (CW) laser through electro-optic modulation at a prescribed RF frequency using EO phase modulators. The comb is then generated at that RF frequency using an EO-intensity modulator. The result is a “minicomb” of spectral bandwidth and number of lines governed by the RF frequency and the number of cascaded phase modulators. The minicomb must then be spectrally broadened to provide coverage over the desired spectrograph bandpass, and spectrally flattened to provide uniform line intensity. The frequency of the pump laser is locked to a stable reference signal to provide the overall comb stability needed for spectrograph calibration. The mode spacing of the \textit{h-, k-} EO astrocomb is 16 GHz, and it provides a comb spectrum from 1.500 $\mu$m to greater than 2.200 $\mu$m. The EO comb is similar in design to the LFC at the Hobby Eberly Telescope’s Habitable Planet Finder \cite{2012SPIE.8446E..1SM}\cite{Metcalf:19}, Palomar’s Radial Velocity Instrument (PARVI) \cite{Gibson_2022}\cite{Cale_2023}, and the Infrared Doppler Instrument on the Subaru Telescope \cite{2018_Subarulfc}. The \textit{h-, k-} astrocomb was delivered to the W.M. Keck Observatory in the summer of 2023 and will be commissioned using the NIRSPEC spectrograph \cite{mcleanDesignDevelopmentNIRSPEC1998} for initial test and validation. An example of the light marker spacing of the HISPEC LFCs is shown in Figure \ref{fig:absrel}.

\subsection{Astro-Etalons}
A set of etalons will be developed to use in tandem with the other calibration sources. This will shield against unreliability and potentially limited lifetime issues with LFCs. An etalon is a resonant cavity that imprints a comb-like spectral response when broadband light is transmitted through it. Etalons can be independently referenced to other ultra-stable sources but are not by default for HISPEC for simplicity. They are typically run at low pressure (sub-mTorr) and are thermally stabilized to $<1$ mK at the zero-expansion temperature of the ultra-low expansion (ULE) material used in the cavity and mirrors ($\sim$30C in most cases).

PARVI uses a similar astro-etalon to the ones specified for HISPEC. The stability of the PARVI Astro-etalon unit was characterized in the laboratory over days. Built by Stable Laser Systems (Boulder, Colorado), they characterized the etalon with a frequency-stabilized laser by scanning the laser over an etalon resonance line and determining the corresponding central wavelength. That central wavelength was then measured over the course of days to determine the drift of the etalon line. The measured drift of the resonance was 17 kHz/day, which corresponds to a radial velocity shift of about 1.7 cm/s/day at 1 $\mu$m, which is acceptable to HISPEC. The drift in frequency is believed to be associated with the ULE relaxing. The ULE will shrink over the course of the first year after it is fabricated at which point the drifts slow.

The broad spectral range of HISPEC (and indeed its bifurcation into two spectrographs) coupled with the technical challenge of manufacturing a serviceable broadband etalon meant HISPEC will make use of two etalons, one for the \textit{y}- and \textit{j}-bands, and the other for the \textit{h}- and \textit{k}-bands.

Other PRV instruments like NEID \cite{NEID_2016_Schwab} and HPF \cite{2012SPIE.8446E..1SM} have used super-continuum sources to illuminate astro-etalons. The benefit of these sources is that they are SMF coupled and have a broad spectrum, both of which agree with the design philosophy of HISPEC. However, using these lamps carries certain drawbacks; the super-continuum source generates on the order of watts of power, and even a highly reflective cavity will take weeks to settle after the light source is fed into the cavity. For the calibration to be useful, therefore, the sources would be turned on full-time to allow the cavity to stabilize. Given the repair cadence and cost of supercontinuum sources, a less costly approach is necessary. Furthermore, after generating several watts of power from the supercontinuum source, the light must then be attenuated achromatically to nanowatts of power to reach flux levels appropriate to be incorporated into science observations with the Hawaii H4RG detector. These pressures led HISPEC to adopt a different option: a tungsten lamp.

These NIR tungsten sources have smooth, near-blackbody spectra generating photons across the entire spectral range of HISPEC. The Thorlabs SLS202L is the source of choice and comes with a condenser lens and an FC/PC optical fiber adapter. We have verified that 300 nW of integrated power can be coupled into an SMF from such a lamp, which will need to be attenuated before it can be routed to the spectrograph, even after being spectrally dispersed at R $= 100,000$. The lamps are predictable, easy, and inexpensive to service over the lifetime of the instrument. 

\begin{figure}[h!]
    \centering
    \includegraphics[width=\linewidth]{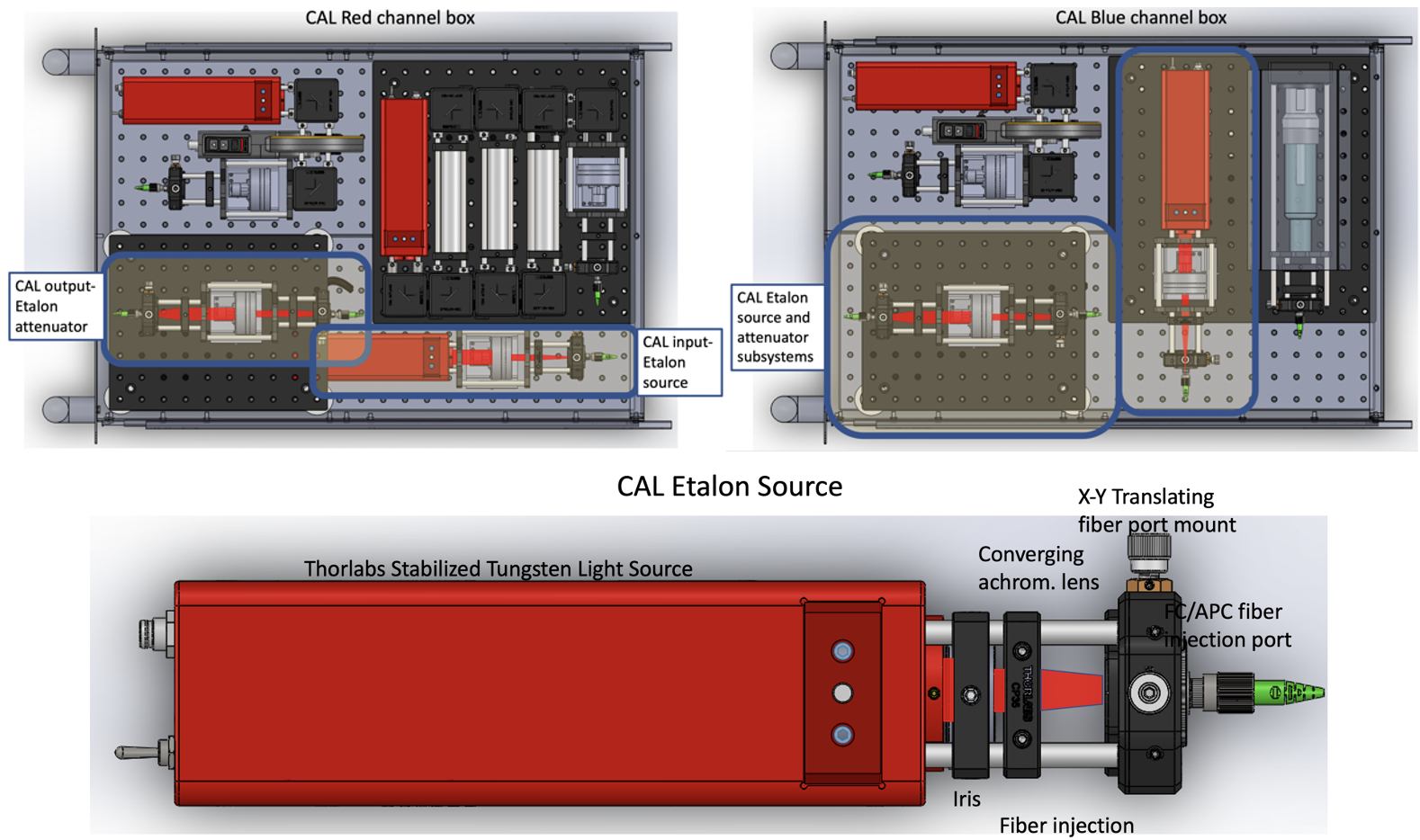}
    \caption{Top left: Cal Red channel box highlighting the red astro-etalon unit. Top right: Cal Blue channel box highlighting the blue astro-etalon unit. Both units utilize a Thorlabs SLS 202L broadband light source and a mechanical, remotely addressable iris to attenuate the light source. Bottom: the injection of the broadband IR light into SMF.}
    \label{fig:etalonbox}
\end{figure}

\begin{figure}[h!]
    \centering
    \includegraphics[width=\linewidth]{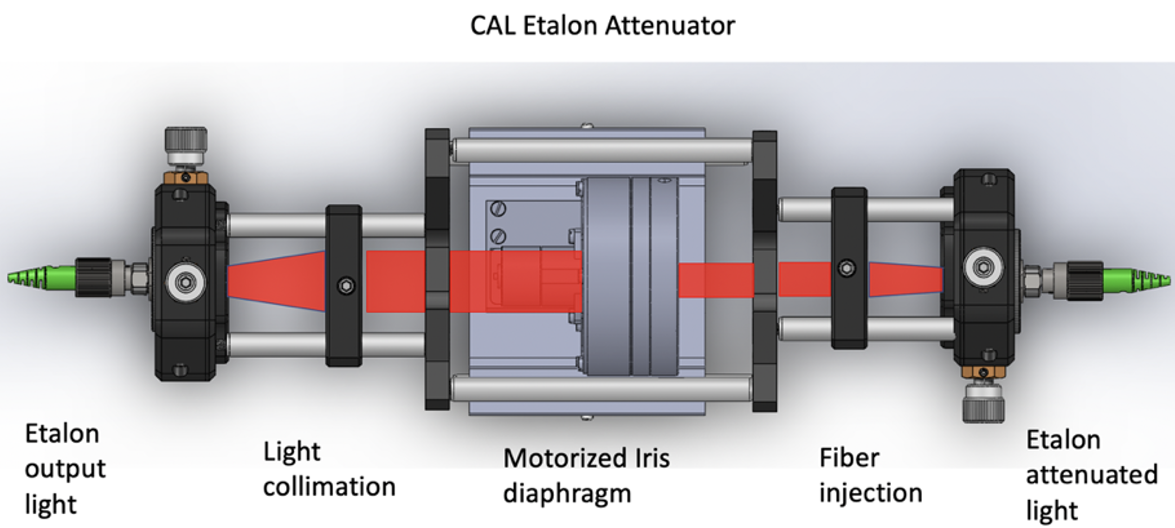}
    \caption{Depiction of the astro-etalon light path through the remotely-addressable motorized iris and instrumentation CAD model}
    \label{fig:etalon_light_path}
\end{figure}

The light source will be hosted in the same box as the flat-field lamps, gas cells, or hollow cathode lamps. It will consist of the lamp, an iris, and an FC/APC optical fiber adapter as shown in Figure \ref{fig:etalonbox}. The beam from the lamp is focused and an SMF collects the light at the focus. Given the source is extended, and there is an abundance of flux, the fiber does not need to be aligned precisely with the beam. The large extent of the beam also makes the fiber alignment highly tolerant and coupled flux is extremely stable over time. A manually adjustable iris will be in the beam before the fiber. The iris will be used to attenuate the light before it goes to the etalon to minimize unnecessary flux being injected into the cavity. The light path of the etalon source is shown in Figure \ref{fig:etalon_light_path}.
\newpage
\section{Discussion}

\subsection{Future Investigation}

\begin{figure}[ht]
    \centering
    \includegraphics[width=\linewidth]{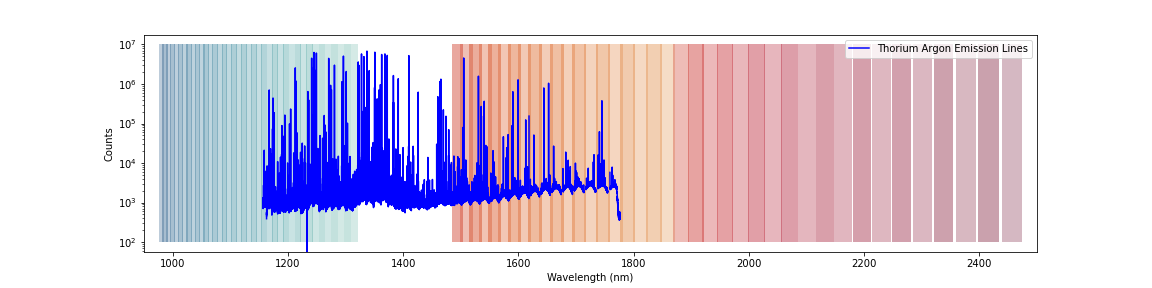}
    \caption{Thorium argon lamp data taken with the PARVI Spectrograph plotted over HISPEC orders. HISPEC BSPEC orders are plotted in varying shades of blue according to wavelength; the same is true for RSPEC. The PARVI bandpass straddles the long side of BSPEC and the short side of RSPEC.}
    \label{fig:ThArPARVIfullbandpass}
\end{figure}

\begin{figure}[ht]
    \centering
    \includegraphics[width=\linewidth]{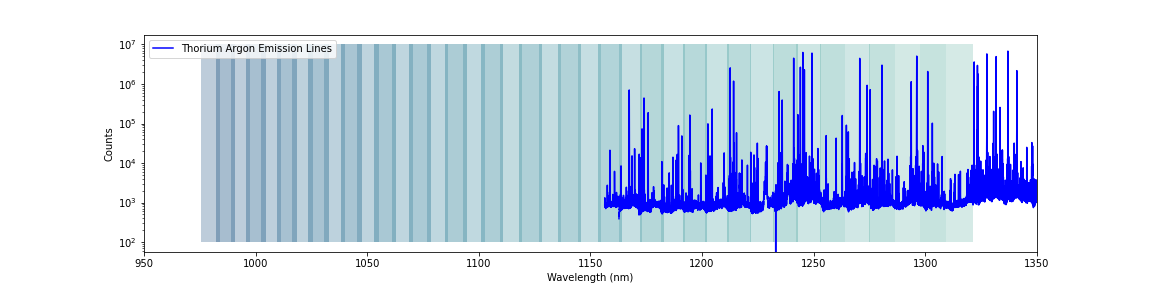}
    \caption{Thorium argon lamp data taken with the PARVI Spectrograph plotted over the BSPEC orders only. Note the PARVI bandpass does not cover the entire spectral range of BSPEC. A high density of emission lines allows for spectral calibration when matching to a Thorium atlas \cite{Redman_2014_Th}. The most intense lines in this data come from the argon fill gas and cannot be used for spectral calibration due to their width and propensity to shift over time.}
    \label{fig:ThArPARVIbluebandpass}
\end{figure}

\begin{figure}[ht]
    \centering
    \includegraphics[width=\linewidth]{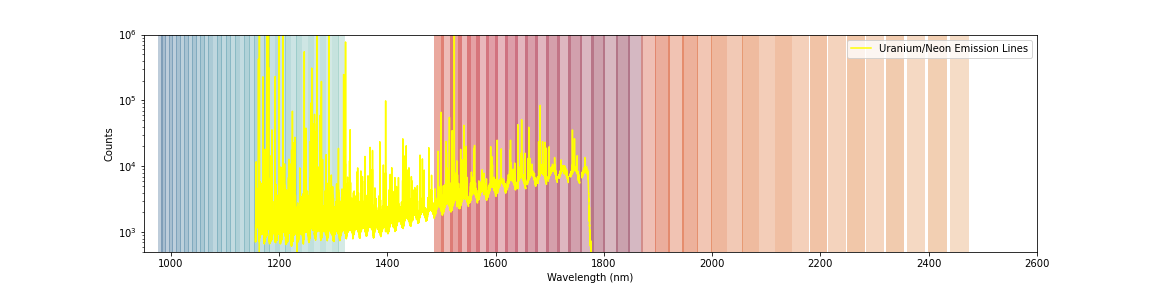}
    \caption{Uranium Neon lamp data taken with the PARVI Spectrograph plotted over HISPEC orders. HISPEC BSPEC orders are plotted in varying shades of blue according to wavelength; the same is true for RSPEC. The PARVI bandpass straddles the long side of BSPEC and the short side of RSPEC.}
    \label{fig:UNePARVIFullbandpass}
\end{figure}

\begin{figure}[ht]
    \centering
    \includegraphics[width=\linewidth]{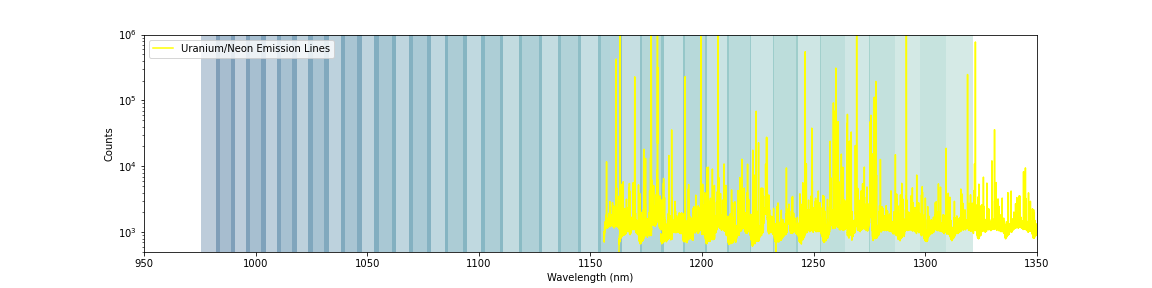}
    \caption{Uranium Neon lamp data taken with the PARVI Spectrograph plotted over the BSPEC orders only. Note the PARVI bandpass does not cover the entire spectral range of BSPEC. A high density of emission lines allows for spectral calibration when matching to a uranium atlas \cite{Redman_2012_uranium}. The most intense lines in this data come from the argon fill gas and cannot be used for spectral calibration due to their width and propensity to shift over time. Work is currently being done to match the data taken with PARVI to uranium line lists. }
    \label{fig:UNePARVIbluebandpass}
\end{figure}

The CAL unit is critical to the successful operation of HISPEC. It must be available whenever HISPEC is on-sky or needs daytime calibrations. As such, it must be robust and reliable. The LFC, etalons, gas cells, and hollow cathode lamps are all technologies that are relatively robust and should suit this application. The implementation of all these sources of calibration provides a level of redundancy in the system that protects against collecting data without wavelength information from calibration. In addition, all the calibration sources act in synergy and can be compared against each other over time to track any drifts in the health of the calibration system and instrument. The calibration mechanical fiber switches should also be robust but are points of failure that need to be carefully monitored and have redundancies built in. Accelerated life testing will be performed in a lab environment to better predict their performance over time.
Before the manufacturing of the CAL subsystem can begin, final decisions about the design must be made regarding our choice of hollow cathode lamp: thorium/argon vs uranium/neon. To validate our choice, we took data with PARVI using both lamps and are in the process of deriving a wavelength solution using this lamp data. With potential ThO contamination threatening the usability of the Th/Ar hollow cathode lamp, our choice of HCL will hinge on which provides the highest density of identifiable emission lines in the BSPEC bandpass of HISPEC. The HCL data taken with PARVI are shown plotted on top of HISPEC orders in Figures \ref{fig:ThArPARVIfullbandpass} and \ref{fig:ThArPARVIbluebandpass} showing the thorium argon HCL supplied by Photron and the uranium neon HCL supplied by Photron in Figures \ref{fig:UNePARVIFullbandpass} and \ref{fig:UNePARVIbluebandpass}. 

To measure subsystem performance, robust laboratory validation will be performed before and during commissioning. The true spectra of the gas cells will be measured at resolutions higher than HISPEC. Long-term laboratory testing will aim to measure any relative drifts based on thermal fluctuations of the etalon and gas cells before and after integration into the total system. Tests will include confirming the drift of the etalon to $\leq$ 20 kHz/day. Testing will be done to determine that the flat-fielding of the detectors meets the requirement of less than 20\% flux variation across the surface of the detector.

\newpage

\acknowledgments 
 HISPEC preliminary design is supported by a gift from the Gordon and Betty Moore Foundation and Caltech.

We acknowledge this instrument is designated to be placed on the summit of Maunakea, a place that has always held a very significant cultural role within the indigenous Hawaiian community. We wish to recognize our privilege to conduct science in this revered location.


\printbibliography 

\end{document}